\documentclass[reprint,superscriptaddress,amsmath,amssymb,aps,pra]{revtex4-2}
 
\usepackage{hyperref}
\usepackage{graphicx}
\usepackage{dcolumn}
\usepackage{bm}
\usepackage{color}
\usepackage{mathtools}
\usepackage{amsmath}  
\usepackage{amssymb} 
\usepackage[thinc]{esdiff}
\newcommand{\re}{{\mathrm e}}

\newcommand{\revision}{\color{black}}
\newcommand{\revisionRef}{\color{black}}

\begin{document}
\preprint{APS/123-QED}
\title{Wave packet dynamics in parabolic optical lattices: From Bloch oscillations to long-range dynamical tunneling}

\author{Usman Ali}
\affiliation{Department of Physics, Paderborn University, Warburger Strasse 100, D-33098 Paderborn, Germany}

\author{Martin Holthaus}

\affiliation{Institut f\"ur Physik, Carl von Ossietzky Universit\"at, 
	D-26111 Oldenburg, Germany}	
	
\author{Torsten Meier}
\affiliation{Department of Physics, Paderborn University, Warburger Strasse 100, D-33098 Paderborn, Germany}
\date{December 27, 2024}    
\begin{abstract}
{We investigate the dynamics of wave packets in a parabolic optical lattice formed by combining an optical lattice with a global parabolic trap. Our study examines the phase space representation of the system's eigenstates by comparing them to the classical phase space of a pendulum, to which the system effectively maps. The analysis reveals that quantum states can exhibit mixed dynamics by straddling the separatrix. A key finding is that the dynamics around the separatrix enables the controlled creation of highly non-classical states, distinguishing them from the classical oscillatory or rotational dynamics of the pendulum. By considering a finite momentum of the initial wave packet, we demonstrate various dynamical regimes. Furthermore, a slight energy mismatch between nearly-degenerate states localized at opposite turning points of the trap potential}  results in controlled long-range dynamical tunneling. These results can be interpreted as quantum beating between a clockwise rotating and a counterclockwise rotating pendulum. 

\end{abstract}

\maketitle
\section{Introduction}
\par{Optical lattices provide a nearly clean environment for the production and manipulation of matter wave packets formed by ultracold atomic ensembles~\cite{1,2,3,4,5}. The advantage is that inter-atomic interactions, environmental decoherence and lattice defects can be tuned as the initial ensemble, the environmental reservoir, and the lattice geometry can be designed properly. The use of Bose-condensed atomic gases in these adjustable lattices translates the complete matter wave picture of de Broglie to spatial scales a hundred thousand times larger and energies at least ten billion times smaller than the usual eV energy scale of solid-state electronic systems~\citep{6,7}. Consequently, a multitude of wave packet phenomena has been realized utilizing this platform, such as superfluidity~\cite{8,9,10}, Bloch oscillations~\cite{11,12}, quantum transport~\cite{13,14}, Anderson localization~\cite{15,16}, Josephson effect~\cite{17,18}, quantum Hall effect~\cite{19,20,21}, and gauge field effects~\cite{22,23}. These phenomena hold great potential for diverse applications in metrology, quantum sensing, imaging, quantum information processing, and computing.}

\par{In optical lattice experiments, parabolic traps are frequently used serving as an auxiliary element for confining and manipulating cold atoms. The parabolic potential aids in confining atoms within a stable region, allowing for a precise control over their spatial distribution. This level of control enables the manipulation of quantum states, the creation of well-defined wave packets, and the simulation of physical systems. The inclusion of a parabolic trap over an optical lattice results in a symmetrically curved periodic lattice, commonly referred to as a parabolic optical lattice~\cite{24}. The combined potential substantially modifies the system's properties compared to systems where only one of the two potentials is present, as shown, e.g.,  in Refs.~\cite{25,NewRef}. This also leads to non-integrability and comprehensive analytical solutions are not possible. However, within the single-band tight-binding approximation one can solve the single-particle system, leading to an analytical description in terms of Mathieu functions~\cite{26,27}. The analytic solutions also predict dipole oscillations of atomic wave packets induced by small displacements of the atomic cloud. In~\cite{28}, the superfluid-dipolar motion is found to be strongly disrupted for large shifts of the parabolic potential, which was first perceived as an insulator-like response. Yet, this phenomenon is identified as the manifestation of Bloch-like dynamics occurring under the influence of a locally-static force of the parabolic potential~\cite{29,30}. A particularly useful description of the parabolic lattice exists in terms of a quantum pendulum model, which delineates the threshold between Bloch and dipole oscillations as the dynamics occurring above and below the separatrix of the pendulum~\citep{29}. 

\par{In this paper, we explore the wave packet dynamics in a parabolic optical lattice by analyzing the energy eigenstates of the system. We consider the tight-binding solutions in which the eigenstates are either parity-related pairs of Wannier-Stark-like localized states lying away from the center of the parabolic lattice or resemble harmonic oscillator eigenfunctions that are localized around the center~\cite{26,27}. We use the phase space representations of the eigenstates and compare them against the classical phase space of the pendulum model. The analysis reveals that the phase space dynamics of states localized away from the trap center and centrally localized states are analogous to the open and closed curves in the pendulum phase space, respectively. Thus, the effectiveness of the previously used quantum pendulum approach is highlighted~\citep{29}. The results also highlight that on and around the separatrix curve there exist numerous states of contrasting nature. Based upon this observation, the dynamics generated by a localized wave packet prepared under the conditions of separatrix are studied. Further, we illustrate different regimes in the system which are tuned by considering a finite momentum for the initial wave packet. Keeping in view the two-fold almost degeneracy maintained by the spatially localized states, a tunneling-like response of localized wave packets is investigated which, unlike the archetypal cases of a step potential~\citep{31,32} or a double well~\citep{33}, is shown to occur even in the absence of a potential barrier. Instead, this type of long-range tunneling appears to be closely related to dynamical tunneling between two separated regions of classical phase space as pioneered by Davis and Heller~\cite{34}, with the distinct difference that in our case the tunneling period can be systematically manipulated to fall within experimentally observable time scales.
\par{The paper is organized as follows: In Section II we introduce the model. We present the important quantum-classical features of the model in Section III, discussing the phase space dynamics. Section IV illustrates the wave packet dynamics corresponding to different regimes of the system. In Section V the long-range tunneling phenomenon is demonstrated. The conclusions along with future perspectives are discussed in Section VI.}

\section{Model} 
\par{Let us start with a single atom in a one-dimensional optical lattice in the presence of a symmetric parabolic trap potential. The Hamiltonian is
\begin{equation} 
	 \hat{H} = \frac{{p}^2}{2m}+\;V_{0}\;\sin^2{\left(\frac{\pi}{a}x\right)}+\frac{1}{2}m\omega^2x^2 \; ,   
	  \label{eq:1}
\end{equation} 
where $\omega$ is the frequency of the parabolic trap, $V_0$ is the optical lattice depth, which is controllable through the intensity of the laser beams, $a$ is the lattice constant, and $m$ is the atomic mass.}

In the limit where the lattice depth significantly exceeds the recoil energy $E_R= (\hbar\pi/a)^2/2m$, whereas simultaneously the change of the parabolic potential over one lattice constant is kept much smaller than the width of the lowest energy band, guaranteeing that the trap-induced tilt of this band exceeds its width only after a large number of sites away from the trap center, the above Hamiltonian is well approximated by the single-band tight-binding model and takes the form
\begin{eqnarray}
	\widehat{H}_{TB} = -J \sum_{n=-\infty}^{\infty}( |n+1 \rangle \langle n| 
        + |n \rangle \langle n+1| ) \notag \\
	  +\Omega \sum_{n=-\infty}^{\infty}  n^2    |n \rangle \langle n| \; , 
	  \label{eq:2}
\end{eqnarray}
where $|n\rangle$ are the ground band Wannier functions, $\Omega = m\omega^2 a^2/2$ represents the strength of the parabolic potential, $J$ denotes the tunneling matrix element which is determined asymptotically by the optical lattice depth, $J \sim \frac{4}{\sqrt\pi} \left(\frac{V_0}{E_R}\right)^{3/4}
\re^{-2\sqrt \frac{V_0}{E_R}}$~\cite{35, 36}.

Let us express the wave function $|\Phi\rangle$ in terms of the Wannier functions as $|\Phi(t)\rangle= \sum_n \phi_n(t)|n\rangle$. The Schr\"odinger equation with the Hamiltonian~\eqref{eq:2} then transforms into the following system of coupled linear equations that govern the time evolution of the complex amplitudes $\phi_n(t)$,
\begin{equation} \label{eq:3}
      i\hbar \dot{\phi}_n 	= -J (\phi_{n+1}+\phi_{n-1}) + \Omega \; n^2 \phi_n. 
\end{equation}
 \begin{figure}[pt]
\centering
\includegraphics[width=0.95 \linewidth]{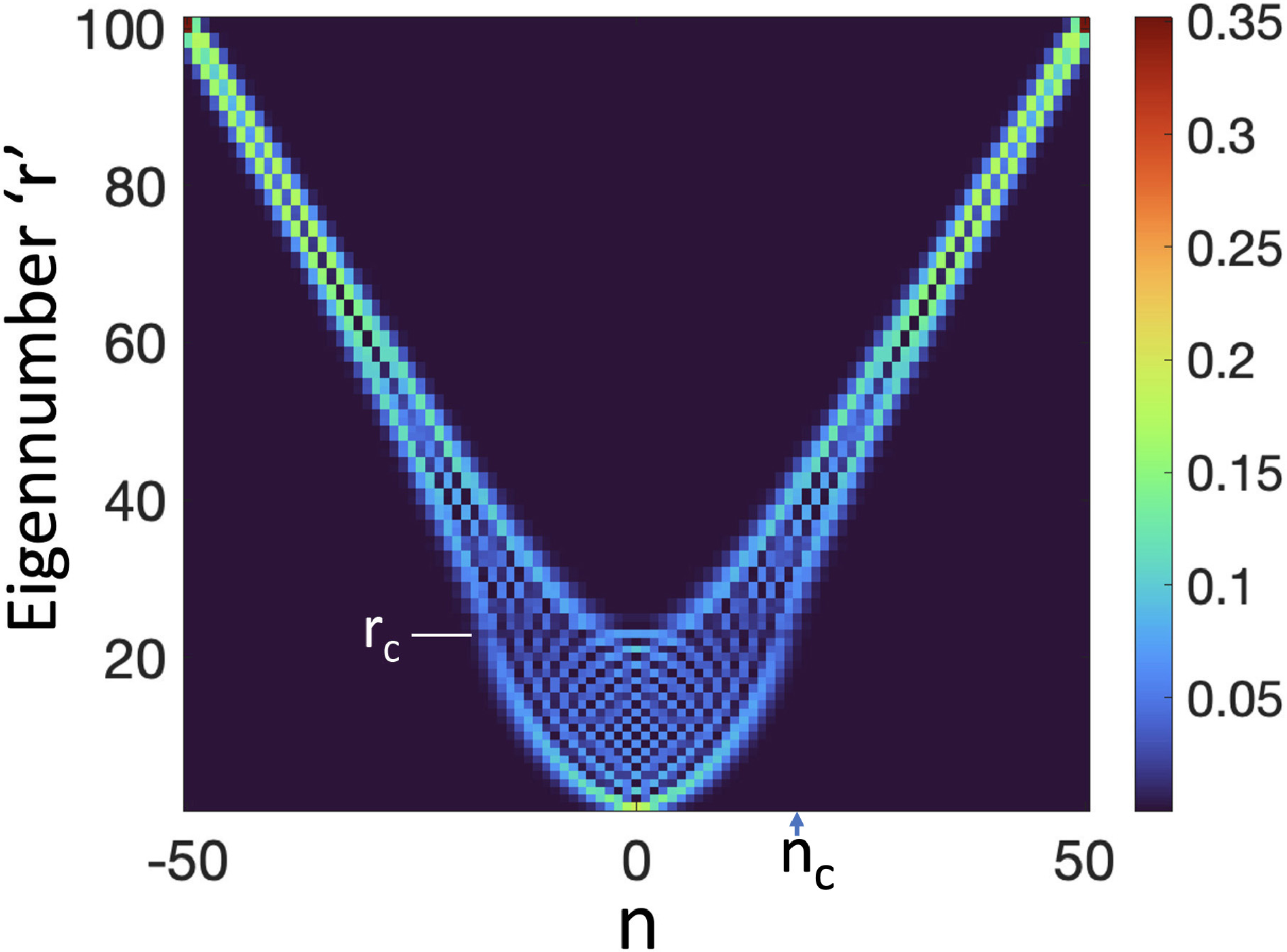}
\caption[eigenstates in parabolic optical lattice]{Absolute squared values of the lowest $100$ eigenstates in Wannier representation, obtained via stationary solutions of Eq.~\eqref{eq:2}. Each eigenstate is offset along the $y$-axis by eigennumber $r$, where $r_c$ marks the critical eigennumber above which the eigenstates change character from harmonic oscillator-like states to Wannier-Stark-like localized states. The parametric values used are: $J= 2.4\times 10^{-2}$~$E_R$ and $\Omega = 3.2\times 10^{-4}$~$E_R$. }
\label{fig:1}
\end{figure}
The above system of equations admits stationary solutions of the form ${\phi^r_n}(t)= {\varphi}^r_n \re^{-iE_rt/\hbar}$, where $\varphi_n^r$ represents the amplitude of the Wannier state associated with the $n$th lattice site for the $r$th eigenstate, and $E_r$ denotes its eigenenergy. Substituting this into Eq.~\eqref{eq:3} results in
\begin{equation} \label{eq:4}
	E_r {{\varphi}}^r_n 	= -J ({\varphi}^r_{n+1}+{\varphi}^r_{n-1}) + \Omega \; n^2 {\varphi}^r_n.
\end{equation}
Representing the stationary amplitudes as the Fourier coefficients of $\pi$-periodic functions $\psi^r(\theta)$, such that}
\begin{equation}	\label{eq:5}
{\varphi}^r_n = \frac{1}{\pi} \int_0^{\pi} d\theta\; \psi^r(\theta) \, \re^{-2in\theta},
\end{equation} 
recasts Eq.~\eqref{eq:4} into a Mathieu equation~\cite{37}
\begin{equation}	\label{Eq:6}	
	\left[\frac{\partial^2}{\partial {\theta}^2}+\left(\frac{4E_r}{\Omega}\right)-2\left(\frac{-4J}{\Omega}\right) \cos{(2{\theta})}\right]{{\psi}^r}(\theta) = 0
\end{equation}
with parameters $\alpha_r = {4E_r}/{\Omega}$ and $q = {4J}/{\Omega}$. The solutions to the above equation are the well-known Mathieu functions, which have been extensively studied and detailed in~\cite{37}. As is well known, the Mathieu equation provides the band edges for a particle in a cosine lattice~\cite{38}. The $\pi$-periodic boundary condition required in the present context simplifies the lattice to a single cosine well, analogous to the potential of a pendulum. The effective Hamiltonian describing the pendulum is 
\begin{figure*}[pt]
\centering
\includegraphics[scale=0.37]{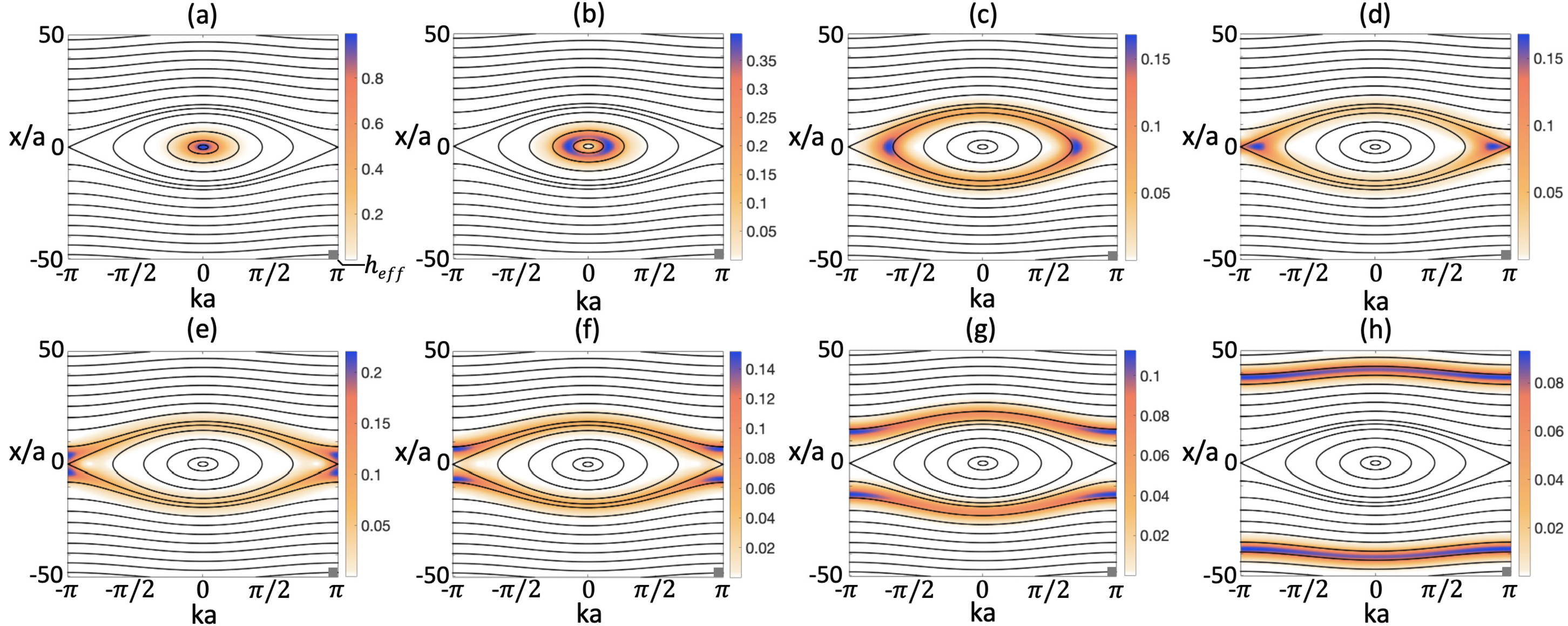}
\caption[eigenstates in parabolic optical lattice]{Husimi distributions for the eigenstates corresponding to $r=0,1,15,20,24,25,35$, and $80$, shown in panels (a) to (h) respectively. These distributions are superimposed on the classical phase space of the pendulum Hamiltonian \eqref{eq:7}. The parametric values remain the same as in Fig.~\ref{fig:1}. Note the gradual shift from the harmonic oscillator-like states aligning with the closed curves to the states localized at positions \(x = \pm {r a}/{2}\) for \(r > r_c\), which evolve according to the open curves. } \label{fig:2} 
\end{figure*}
\begin{equation} \label{eq:7}
  \hat{\mathcal{H}}= \frac{\Omega}{4} \hat{L}^2\;-2J\cos(2\theta),  \quad \hat{L} = -i\frac{\partial}{\partial \theta},
\end{equation}
with $\hat{L}$ denoting the angular momentum of the pendulum. Hence, the tight-binding system \eqref{eq:2} maps onto the pendulum model, with the clear advantage of interpretation in terms of pendulum dynamics. Also, Hamiltonian \eqref{eq:7} directly corresponds to Eq.~\eqref{eq:2} when expressed in the Bloch basis. In this context, the angular momentum \(L\) is related to the spatial position as \(L = {2x}/{a}\), and the angular position \(\theta\) of the pendulum is connected to the atomic quasimomentum \(k\) through the relation \(\theta = {ka}/{2}\). 
 
\par{While the Mathieu functions provide an analytical representation of the eigenstates~\cite{26}, instead of delving into their specifics we numerically diagonalize Eq.~\eqref{eq:2} to obtain the stationary eigenstates. The parameter values employed in this work correspond to an experiment involving $\textsuperscript{87}$Rb atoms in an optical lattice with a depth of $V_0=10$~$E_R$ and a lattice constant of $a=397.5$~nm, supplemented by a parabolic trap with a trapping frequency $\omega=2 \pi \times 36$~Hz. These data amount to the parameters $J= 2.4\times 10^{-2}$~$E_R$, $\Omega = 3.2\times 10^{-4}$~$E_R$, and $q=300$. Here the Mathieu parameter $q = 4J/\Omega$ obtains further intuitive significance: The energy shift $\Omega n^2$ induced by the trap matches the band width $4J$ after about $\sqrt{q}$ lattice sites, implying that the condition $\sqrt{q} \gg 1$ secures the validity of the single-band approximation~\eqref{eq:2}}.}
\par {Figure~\ref{fig:1} displays the numerically calculated lowest $100$ eigenstates of the system, plotted in Wannier representation. The lower-lying states are harmonic oscillator-like localized around the trap center, which are followed by states that are increasingly localized into two separate regions of space, akin to Wannier-Stark localization in a locally-linear potential. The distinction lies in the energy of states being below or above the band edge $2J$ of the periodic lattice. Physically, this depends upon whether the tunneling or the trapping strength is playing a more significant role. The solutions dominated by tunneling then correspond to the tight binding regime, while the solutions dominated by the trap potential itself belong to the weak binding regime. The critical eigennumber separating these regimes is approximately given by $r_c=||\sqrt{2q}||$, where $||y||$ denotes the integer nearest to $y$. The states above $r_c$ correspond to nearly degenerate pairs with opposite parity centered around each point $n = \pm r/2$. These exhibit a minuscule energy splitting approximately on the order of $q^r/r^{(r-1)}$, where $r\gg\sqrt{q}$~\cite{37}. As a result, they are connected across large spatial distances by quantum tunneling.}

\section{Eigenstates in phase space}
\par{To develop insights into the wave packet dynamics, it is valuable to compare the classical phase-space structure of the pendulum with the phase space representation of eigenstates. The Husimi representation~\cite{39} provides an effective way to visualize a wave function and enables a direct comparison with the classical phase space. The Husimi Q-function is constructed by taking the squared projection
\begin{equation}
	Q_{\chi}(x,k) =	|\langle \alpha_{x,k}|\chi\rangle|^2
\end{equation}
 of the wave function $|\chi\rangle$ with the coherent state $|\alpha_{x,k}\rangle$ peaked at coordinates $(x,k)$ in the phase space. The coherent state in real-space representation is expressed as}
\begin{equation}
    \alpha_{x,k}(x')=\langle x'|\alpha_{x,k}\rangle	= \frac{1}{\sqrt{\sigma_{x} \sqrt{\pi}}} \; \re^{-\frac{(x'-x)^2}{2\sigma_{x}^2}} \re^{-ik{(x'-x)}}.
\end{equation}
\par{In Fig.~\ref{fig:2} we show the Husimi distributions obtained for specific instances of the eigenstates at $r=0,1,15,20,24,25,35$, and $80$, that are superimposed on the classical phase space corresponding to the pendulum Hamiltonian \eqref{eq:7}. In terms of classical dynamics, the Hamiltonian \eqref{eq:7} refers to the closed and open curves in phase space which represent the vibrational and rotational regimes of the pendulum~\cite{40}. These regimes are separated by a specific curve known as the separatrix, which is determined by the relation
\begin{equation}  \label{eq:10}
	x_c = a\sqrt{\frac{2J}{\Omega}[1+\cos(ka)]}
\end{equation}
between position and quasimomentum.

Figure~\ref{fig:2}(a) and (b) showcase the ground and the first excited state, $r=0$ and $1$, clinging to the closed curves below the separatrix. This depicts the harmonic oscillator-like character of the low-energy states, where the period of oscillations is $T_D=\pi\hbar/\sqrt{J\Omega}$. For $r\sim \sqrt{2q}$, the harmonic oscillator-like states become closer to the separatrix and the intermediate states start to appear that propagate spreading along the separatrix curve. This behavior is highlighted in Fig.~\ref{fig:2}(c) and (d) for $r=15$ and $20$. The states start to localize in two separate regions of space at $r\sim r_c$, which becomes completely apparent for $r > r_c$,  and the localized densities evolve as per the open curves, as shown in Fig.~\ref{fig:2}(e) and (f) for $r=24$ and $25$. It should be noted that the overall density for states slightly above $r_c$ still remain on and around the separatrix and it deviates from the separatrix curve near the boundaries of the Brillouin zone. Thus, on and around the separatrix there exist three type of states, which we will later show in our analysis to give rise to highly nonclassical dynamics}. The Wannier-Stark-like localized states, which fully adhere to the open curves, are also shown in Fig.~\ref{fig:2}(g) and (h). The localization of these states is also discussed in reference~\cite{25} employing a semiclassical viewpoint which establishes a connection between the emergence of new turning points due to lattice-induced Bragg scattering and the classical turning points of the parabolic trap potential. In this regard, the phase space dynamics found in our analysis accentuate the oscillations between the turning points which are in agreement with the previously developed approximate theoretical description. Because the system possesses reflection symmetry with respect to the trap center, its eigenstates alternatingly have even or odd parity, such that states above the separatrix show up as almost degenerate pairs with opposite parity. The situation encountered here is similar to the case of a symmetric double well where pairs of eigenstates with opposite parity appear for energies below the barrier, such that their even or odd linear combinations are localized in only one of the wells. In our case there is no tunneling through a barrier, but quantum tunneling through a classically forbidden region of phase space instead~\cite{34}, such that each member of a parity-related pair is localized around both of its turning points, whereas their even or odd linear combinations are localized at one of these points only. It is this feature which enables the long-range dynamical tunneling effect which we will explore in detail in Section~V.
 \begin{figure}[pt]
\centering
\includegraphics[width=0.95 \linewidth]{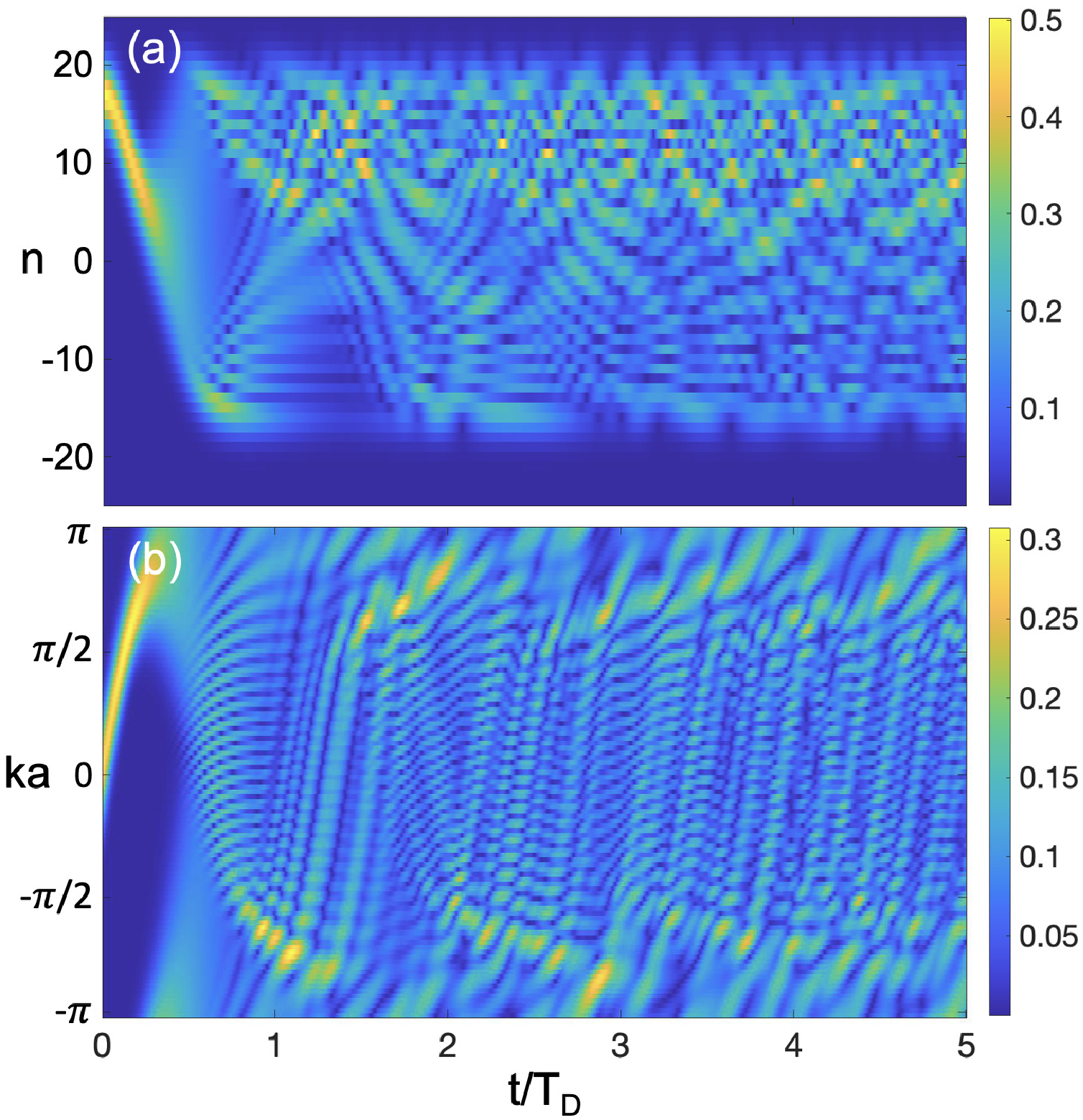}
\caption[eigenstates in parabolic optical lattice]{Dynamics corresponding to the separatrix of the quantum pendulum. Absolute value of the wave packet evolution in real- and quasimomentum-space is shown in (a) and (b), respectively. The initial wave packet is placed at $n_0=17$ with $k_0a=0$ and $\sigma_0=2.23$. The parametric values used are $J= 2.4\times 10^{-2}$ $E_R$ and $\Omega = 3.2\times 10^{-4}$ $E_R$ which are same as in previous figures. See the Supplemental Material \cite{NewRefSM} for a movie illustrating the dynamics in the Husimi representation within the pendulum phase space.}
\label{fig:3}
\end{figure}
 \begin{figure}[pt]
\centering
\includegraphics[width=0.96 \linewidth]{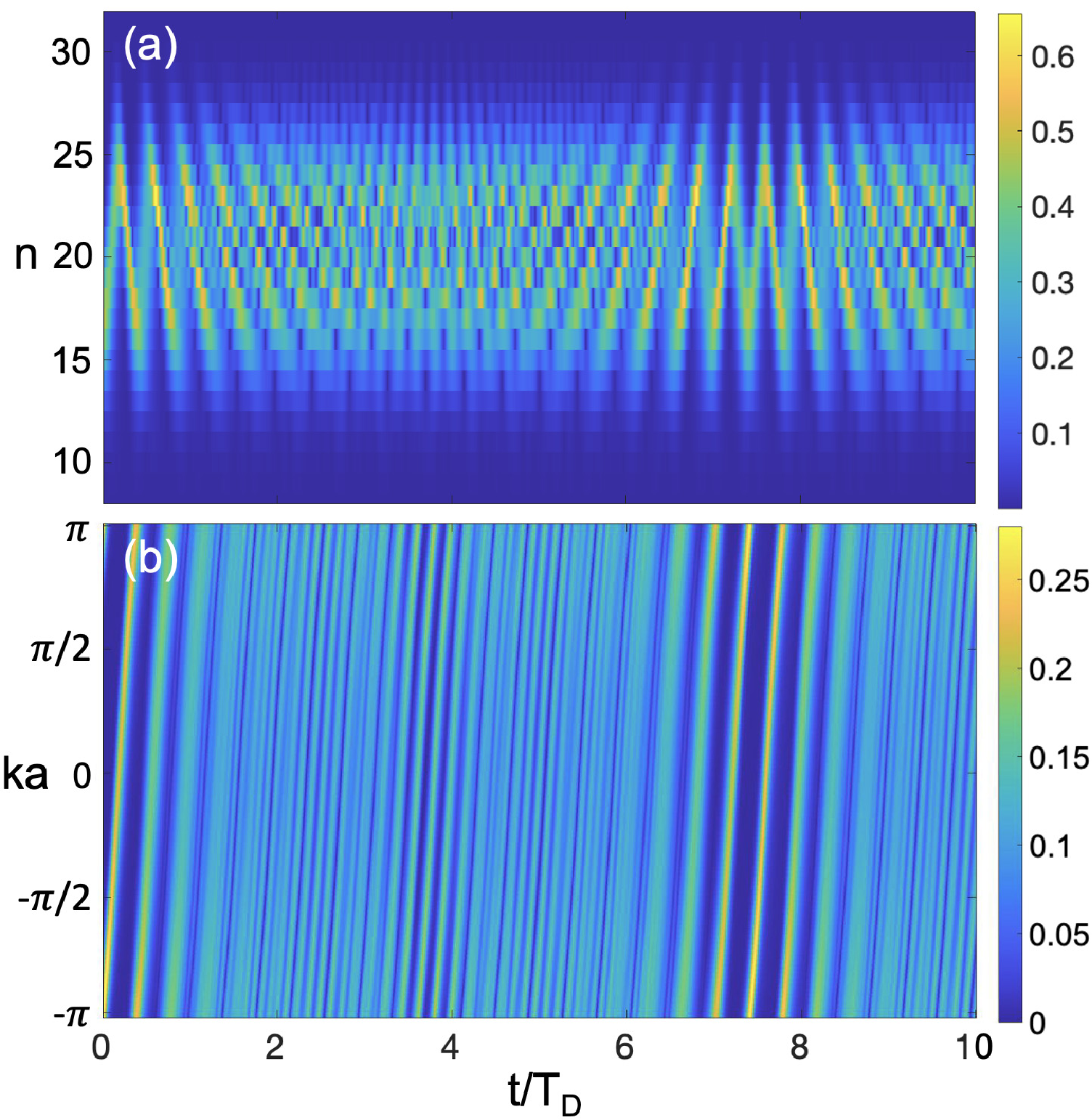}
\caption[eigenstates in parabolic optical lattice]{Bloch-like oscillations above the separatrix. Absolute value of the wave packet evolution in real- and quasimomentum-space is shown in (a) and (b), respectively. All the parametric values remain the same as in Fig.~\ref{fig:3}, changing only the initial quasimomentum to $k_0a=\pi$.}
\label{fig:4}
\end{figure}
\section{Near-Separatrix dynamics}
\par{Next, we choose an initial wave packet and demonstrate the dynamics by placing it in different regions of space around the separatrix. The initial wave packet is taken as a localized Gaussian, 
\begin{equation} 
	|\Psi_0\rangle = \sum_n \frac{1}{\sqrt{\sigma_0 \sqrt{\pi}}}\; e^{-\frac{(n-n_o)^2}{2\sigma_0^2}} e^{-ik_0a{(n-n_o)}}|n\rangle,	
\end{equation}
where, $n_0$, $k_0$, and $\sigma_0$ represent the initial mean position, quasimomemtum, and width of the wave packet. The time evolution of the wave packet is obtained by solving the Eq.~\eqref{eq:3}, and the quasimomentum space dynamics are plotted by taking the Fourier transform of the real-space evolution.}
\par{Figure~\ref{fig:3} displays the dynamics generated by the wave packet placed at the spatial location equivalent to the separatrix at zero quasimomentum, i.e. $n_c=17$. Clearly, the wave packet performs mixed dynamics which is due to the presence of three distinct types of states around the separatrix: harmonic oscillator-like states, intermediate states, and Wannier-Stark-like localized states. This is confirmed by the wave packet evolution in real and quasimomentum space shown in Fig.~\ref{fig:3}(a) and (b), respectively. The wave packet spreads at the time corresponding to half of the Bloch period where a fraction of the total density oscillates around the center of the Brillouin zone, while the remaining part undergoes Bragg reflection. In terms of pendulum dynamics at the separatrix, this is equivalent to the wave packet first moving towards the hyperbolic fixed point and then splitting, with one part corresponding to a clockwise rotating pendulum and the other to a counterclockwise rotating pendulum, leading to subsequent multiple interferences (see the movie in the Supplemental Material~\cite{NewRefSM}). Thus, the wave packet dynamics is a mix of harmonic oscillator-like and Bloch oscillation-like dynamics. Consequently, the wave packet spreads over the entire range which is energetically accessible, still preferentially populating the wing of the parabolic trap it was initially prepared in.}
 \begin{figure}[pt]
\centering
\includegraphics[width=0.94 \linewidth]{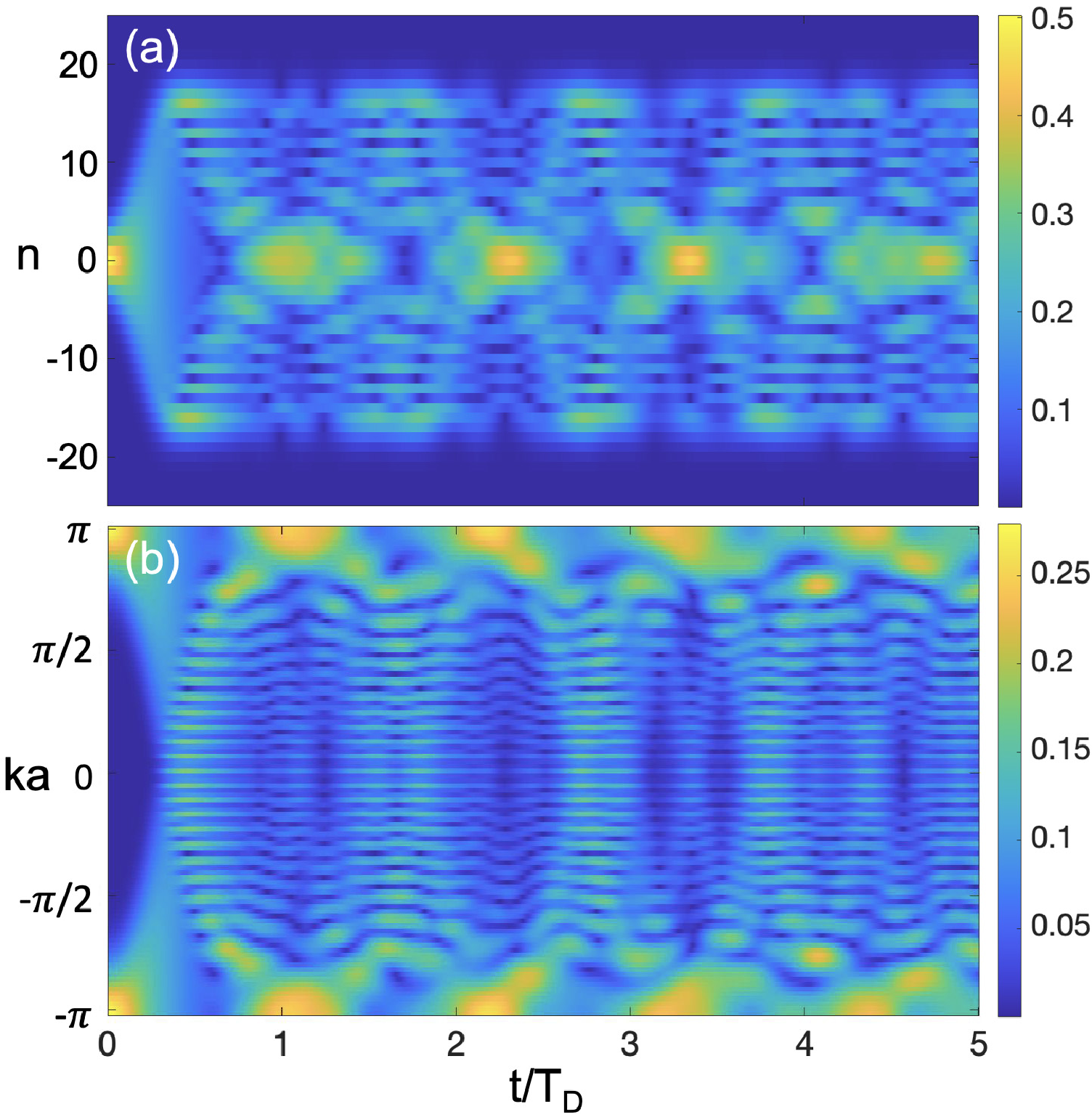}
\caption[eigenstates in parabolic optical lattice]{Wave packet dynamics along the separatrix curve. Absolute value of the wave packet evolution in real- and quasimomentum-space is shown in (a) and (b), respectively. All the parametric values remain the same as in Fig.~\ref{fig:3}, with only the initial position changed to \(n_0 = 0\) and the initial quasimomentum to \(k_0 a = \pi\). See the Supplemental Material \cite{NewRefSM} for a movie showing the Husimi distribution of the dynamics in the pendulum phase space.}
\label{fig:5}
\end{figure}
\par{Notably, the time evolution of the mean position in the above dynamics decays toward the origin within the first period. Such a decay has been observed in experiment~\cite{28} and is attributed to an inhibition of oscillations. A more comprehensive understanding can be gained by examining the overall evolution of the wave packet, rather than focusing solely on the expectation values. Our analysis reveals the presence of mixed dynamics, offering insights into the behavior near the separatrix.}
 \begin{figure}[pt]
\centering
\includegraphics[width=0.935 \linewidth]{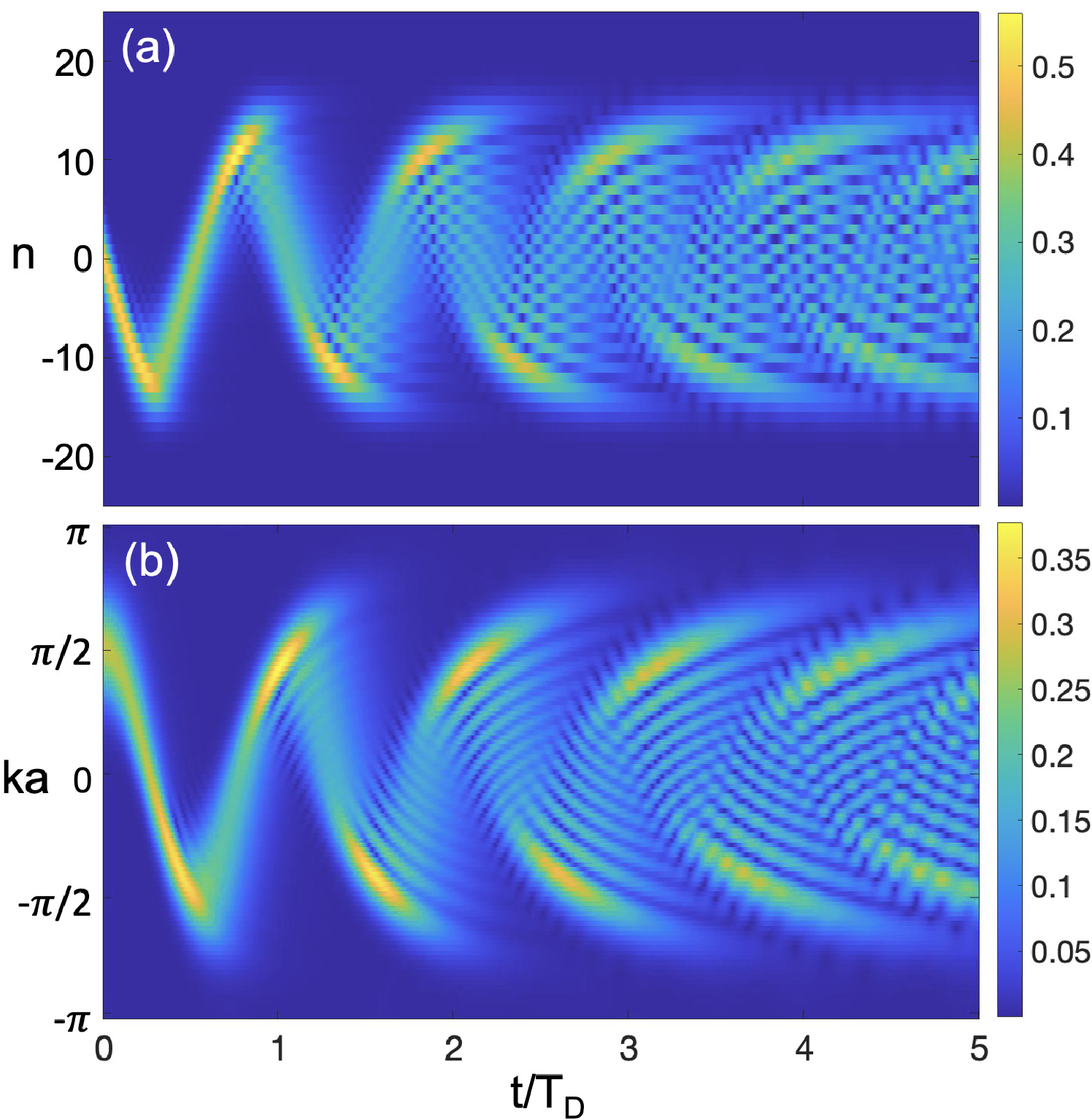}
\caption[eigenstates in parabolic optical lattice]{Dipole oscillations below the separatrix. Absolute value of the wave packet evolution in real- and quasimomentum-space is shown in (a) and (b), respectively. All the parametric values remain the same as in Fig.~\ref{fig:3}, with only the initial position changed to \(n_0 = 0\) and the initial quasimomentum to \(k_0 a = \pi/2\). See the Supplemental Material \cite{NewRefSM} for a movie showing the Husimi distribution of the dynamics in the pendulum phase space.}
\label{fig:6}
\end{figure}
\par{Further, considering the above phase space analysis, we note that for the initial wave packet placed in the vicinity of the separatrix, the dynamical evolution depends strongly on the initial momentum of the wave packet. Thus, for different choices of \(n_0\), we vary \(k_0 a\), which allows us to tune the dynamics.}
\par{The wave packet dynamics with $n_0=17$ and $k_0a= \pi$ are shown in Fig.~\ref{fig:4} which reveal a transition from mixed dynamics to Bloch-like oscillations induced by the momentum shift. For $k_0a= -\pi$ or $\pi$, the separatrix curve reaches the origin, such that for $n_0=17$ the dynamics corresponds to the open curves. The wave packet oscillations in a restricted region of space on one side of the parabolic lattice are visible in Fig.~\ref{fig:4}(a). These oscillations alongside the Bragg reflections in the quasimomentum-space dynamics shown in Fig.~\ref{fig:4}(b) confirm the associated Bloch-like dynamics. The Bloch-like oscillations dephase quite quickly which is due to the unequal energy spacing between lattice wells. The dephasing is followed by periodic revivals, as shown in Fig.~\ref{fig:4}.}
\par{Furthermore, we take the wave packet to the origin, i.e., $n_0 = 0 $. In this case, the wave packet would weakly breathe and is expected to remain confined to the center of the parabolic lattice when $ k_0a = 0 $ (not shown). However, for $ k_0a = \pi $, the dynamics become much more intriguing. The wave packet spreads along the separatrix, with partial reconfinement occurring over time in the presence of small in-well oscillations, as shown in Fig.~\ref{fig:5}. This is because the wave packet is initially placed precisely on the unstable hyperbolic fixed point. In the classical scenario, a pendulum at the hyperbolic fixed point would break symmetry and fall either to the right or left when subjected to an infinitesimally small perturbation. However, the quantum system retains the initial symmetry, and both paths are followed simultaneously (see the movie in the Supplemental Material~\cite{NewRefSM}). This highlights the non-classical modification of the dynamics in this case.
 \begin{figure}[pt]
\centering
\includegraphics[width=0.94 \linewidth]{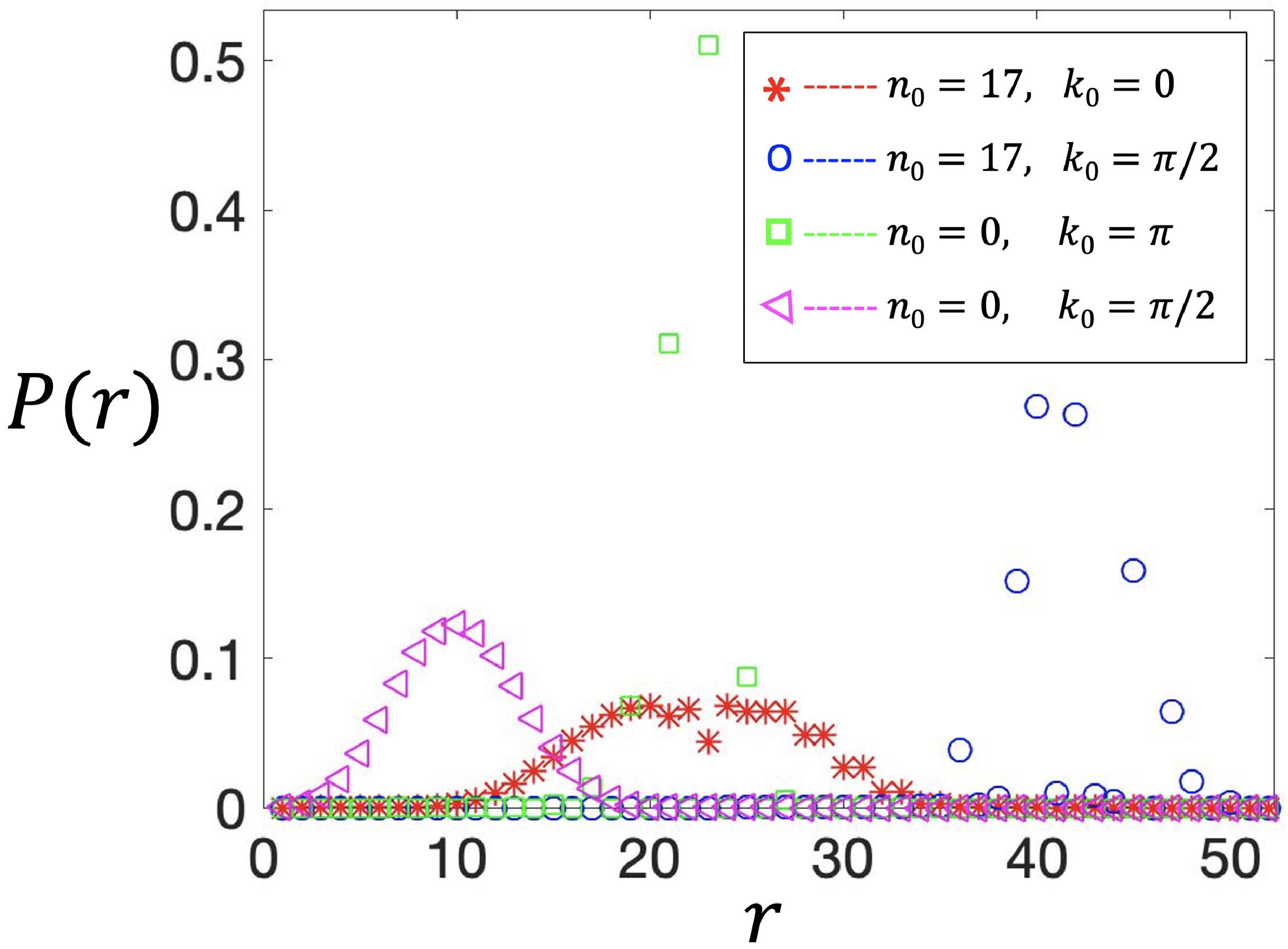}
\caption[eigenstates in parabolic optical lattice]{Occupation probabilities of the eigenstates $| \Phi^r \rangle$, as provided by the squared overlap $| \langle \Psi_0 | \Phi^r \rangle|^2$. The state $|\Psi_0\rangle$ corresponds to the initial Gaussian wave packet considered in the different scenarios in Figs.~\ref{fig:3}-\ref{fig:6}. The attribution of the symbols to these cases can be inferred from the legend.}
\label{fig:7}
\end{figure}

On choosing $ k_0a = \pi/2$ with $n_0 = 0 $, the wave packet performs dipole oscillations. Nearly harmonic oscillations across the center of the parabolic lattice are clearly visible in Fig.~\ref{fig:6}(a). The oscillations of the quasimomentum around the center of the Brillouin zone are also evident from the quasimomentum-space dynamics shown in Fig.~\ref{fig:6}(b). Similar to Bloch-like dynamics discussed above, the dipole oscillations also carry an intrinsic dephasing, leading to a collapse of the oscillatory dynamics, followed by subsequent revivals.

In order to distinguish between the eigenstates of different nature which are populated by the Gaussian wave packet in each of the various scenarios considered in Figs.~\ref{fig:3} to~\ref{fig:6}, we present in Fig.~\ref{fig:7} a comparison between occupation probabilities of eigenstates obtained in each case. The occupation probabilities are plotted through the absolute-squared projection of the initial Gaussian wave packet on the eigenstates. The results highlight that for the wave packet prepared under the conditions of the separatrix with $ n_0 = 17$ and $ k_0a = 0$, as in Fig.~\ref{fig:3}, approximately 22 states are populated, exhibiting a mixed population of harmonic oscillator-like, intermediate, and localized Wannier-Stark-like states. This reflects the mixed dynamics in this case. In contrast, for the Gaussian wave packet prepared well above the separatrix with $ n_0 = 17$ and $ k_0a = \pi$, such as considered in Fig.~\ref{fig:4}, the population is entirely on localized Wannier-Stark-like states, indicating Bloch-like oscillations. Further, the wave packet taken to $n_0 = 0$ and $ k_0a = \pi$ reveals a concentration of population in four intermediate states and two localized states, where the state at quantum number $r=23$ carries more than half of the occupation. These states have all been identified in Sec~II as evolving along the separatrix, which reaffirm the breathing dynamics in Fig.~\ref{fig:5}. Next, in the case with $n_0 = 0$ and $ k_0a = \pi/2$ the occupation of harmonic oscillator-like states underscores the dipole oscillations reported in Fig.~\ref{fig:6}. This analysis sheds light on the different types of dynamical behavior based upon the nature of eigenstates.
}
\section{Quantum tunneling above the separatrix}
\par{For significantly larger shifts of the initial wave packet with respect to the center of the parabolic trap, the multi-band structure of the full system~\eqref{eq:1} undermines the single-band tight-binding approximation~\eqref{eq:2}. Namely, when the quadratic tilt $\Omega n^2$ bridges the energy gap~$\Delta$ between the lowest and the first excited energy band, that is, for initial shifts $n_0 > n_{\rm max} = \sqrt{\Omega/\Delta}$, Landau-Zener tunneling between these two bands sets in~\cite{29,41}; note that $n_{\rm max} = 129$ for the parameters adopted in the present work. In this case the tunneled fraction of the wave packet again would undergo harmonic oscillator-like dynamics in the upper band, such that the overall dynamics turn into an intricate two-band superposition of Bloch-like and harmonic oscillator-like oscillations. In the present investigation, however, we do not consider such interband tunneling, but tunneling between different regions of space corresponding to the same local tilt of the lowest band with opposite sign. This is achieved by restricting ourselves to initial shifts which obey the condition $n_c < n_0 <n_{\rm max}$.}       
\begin{figure}[pt]
\centering
\includegraphics[width=0.85 \linewidth]{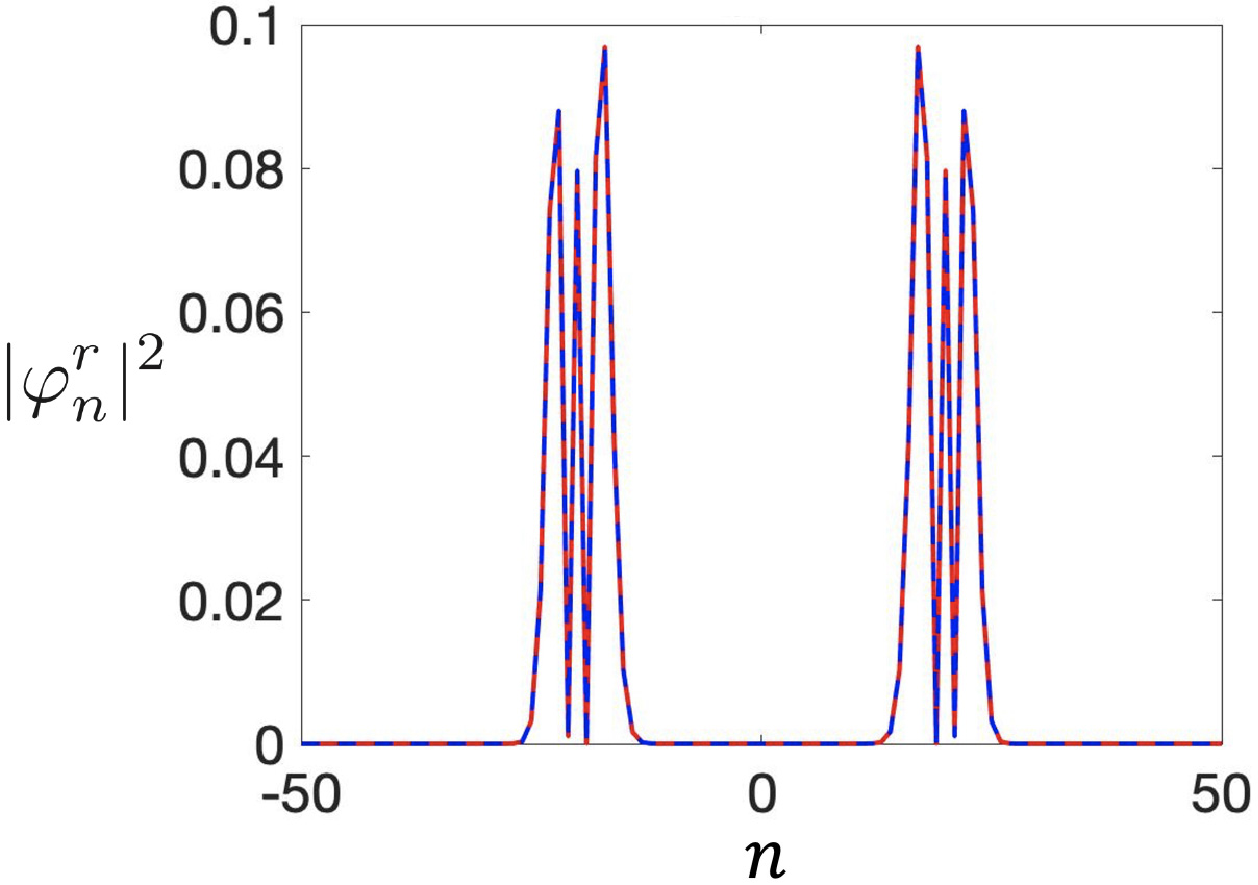}
\caption[eigenstates in parabolic optical lattice]{Absolute-squared values of the wave function amplitudes of a nearly-degenerate pair of states as a function of $n$ in the presence of an additional energy shift. The state at \( r = 40 \) (red dashed curves) and \( r = 41 \) (blue dashed curves) are depicted with \( \epsilon = 3.6 \times 10^{-4} E_R \). All other parameters are the same as in Fig.~\ref{fig:1}.}
\label{fig:8}
\end{figure}
\par{As already indicated at the end of Sec.~III, the tunneling process we are interested in emerges from the nearly-degenerate eigenstates above the separatrix. A transition of quantum particles across nearly-degenerate states is a fascinating aspect of quantum tunneling~\cite{31,32,33,42,43}. This behavior arises due to the non-zero probability of finding the particle in a classically forbidden region, governed by the wave function's exponential decay beyond a potential barrier. The tunneling time is determined by the energy difference $\Delta E$ between symmetry-related states according to~\cite{44}
\begin{equation} \label{eq:12}
	T_{\rm tun} = \frac{\pi\hbar}{\Delta E} \; .
\end{equation}
This is the time during which a wave packet completely transfers from one state to the other. Thus, if $\Delta E$ is very small, the tunneling occurs on large time scales. Keeping in view the minuscule energy splitting between the nearly-degenerate states in the model considered above, the corresponding tunneling times would be much larger than the current experimentally measurable time scales in cold atom experiments, which range from a few hundred milliseconds (ms) to a few seconds.}
\par{However, there exists an intriguing and experimentally feasible way to strongly reduce the long-range tunneling times in parabolic optical lattices. Namely, if one adds a second, much weaker lattice with twice the period of the primary lattice, one introduces an energy mismatch $\epsilon$ between neighboring sites, as described by the modified tight-binding Hamiltonian} 
\begin{eqnarray} \label{eq:13}
	\widehat{H}' = -J \sum_{n=-\infty}^{\infty}( |n+1 \rangle \langle n| 
        + |n \rangle \langle n+1| ) \notag\\
	+  \sum_{n=-\infty}^{\infty} \big(\Omega n^2  + \frac{\epsilon}{2} \:(-1)^n \big) |n \rangle \langle n|\;.
\label{eq:19}
\end{eqnarray}
\par{If the energy mismatch is on the order of the hopping matrix element $J$, the binary lattice dimerizes and therefore possesses two Bloch bands, offering one of the simplest setups for investigating interband tunneling effects~\cite{45}.  Accordingly, the Landau-Zener interband tunneling dynamics, effectuated by an external constant force, have been studied in such systems~\cite{46}. In contrast, here we consider a binary parabolic lattice with a very small mismatch $\epsilon \ll J$, such that the lattice effectively can still be described by a single band.}
\begin{figure}[pt]
\centering
\includegraphics[width=0.93 \linewidth]{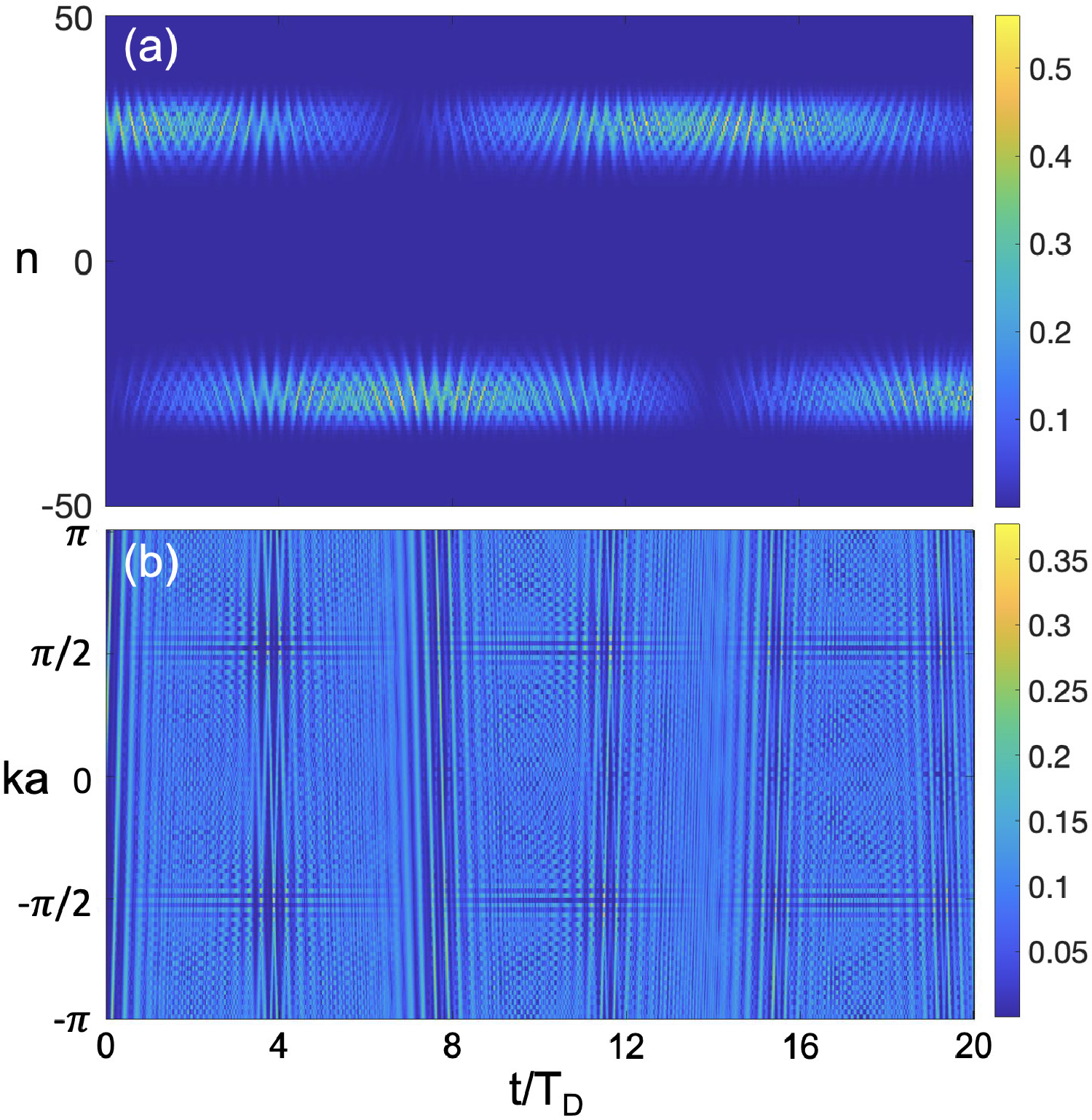}
\caption[eigenstates in parabolic optical lattice]{Quantum tunneling atop Bloch oscillations. Absolute value of the wave packet evolution in real- and quasimomentum-space is shown in (a) and (b), respectively. All the parametric values remain the same as in Fig.~\ref{fig:5}, changing only the initial position $n_0= 30$ and the initial quasimomentum $k_0a=0$, and choosing $\epsilon=3.6 \times 10^{-4}E_R$. The tunneling time is related to the dipole period by $T_{tun}=7.18T_D$.}
\label{fig:9}
\end{figure}
In that case the previous Eq.~\eqref{eq:12} is replaced approximately by the expression
\begin{equation} \label{eq:14}
T_{\rm tun} \approx \frac{\pi \hbar}{\epsilon} \; ,
\end{equation}
implying that the strength of the binary lattice allows one to tune the tunneling time. While the secondary lattice thus increases the  tunneling splitting, it does not affect the presence of symmetry-related pairs of eigenstates, which is a precondition for the tunneling effect to occur. This is confirmed by Fig.~\ref{fig:8}, where we present the absolute-squared values of the wave function amplitudes for such a pair. Thus, the even or odd linear combinations of each symmetry-related pair actually are located on the right or left wing of the parabola, respectively, which is what enables long-range tunneling between these arms.
\begin{figure}[pt]
\centering
\includegraphics[width=1. \linewidth]{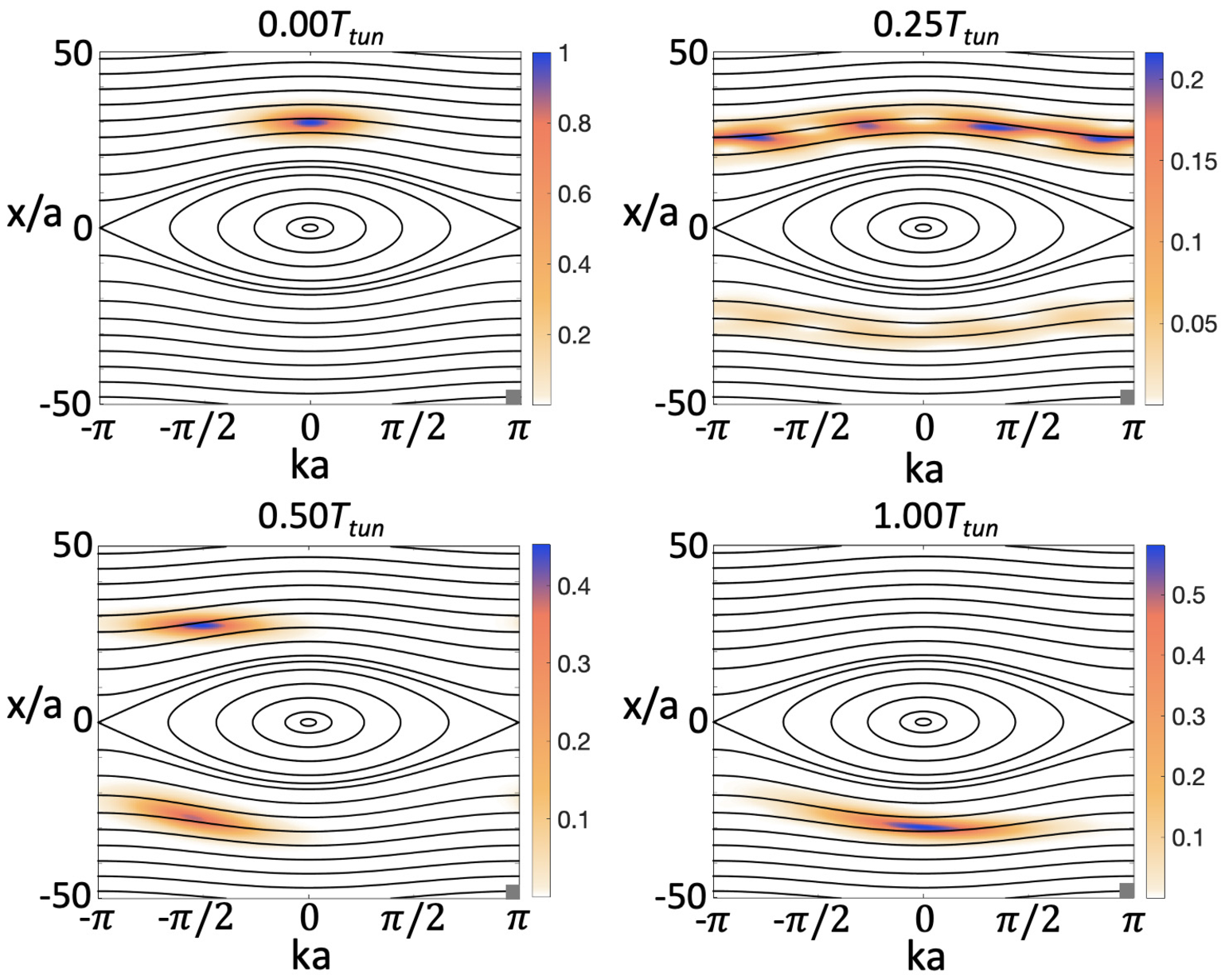}
\caption[eigenstates in parabolic optical lattice]{Evolution of the Husimi distribution of Gaussian wave packet tunneling in the pendulum phase space. Snapshots of the simulated Husimi distribution for the wave packet evolution shown in Fig.~\ref{fig:9} are presented. The dynamics are illustrated for one tunneling period, $T_{tun}=7.18T_D$. A movie showing the evolution of the Husimi distribution is provided in the Supplemental Material~\cite{NewRefSM}.}
\label{fig:10}
\end{figure}
The tunneling effect is demonstrated in Fig.~\ref{fig:9}, where we suppose the Gaussian wave packet at $n_0=30$ with $k_0a=0$, therefore it is kept well above the separatrix. Accordingly, the wave packet initially performs Bloch oscillations on one wing of the parabolic lattice, on top of which it also tunnels and eventually appears on the other arm. This tunneling process is gradual and repeats itself continuously. Thus, the wave packet moves back and forth across both the arms, maintaining coherence across large distances, see Fig.~\ref{fig:9}(a). In fair agreement with Eq.~\eqref{eq:14} the tunneling time numerically observed for $\epsilon = 3.6 \times 10^{-4} E_R$ is close to $7.18$ dipole periods, which is equivalent to nearly $357$~ms. Moreover, the $k$-space dynamics shown in Fig.~\ref{fig:9}(b) highlights the inverted momenta of the tunneled wave packet due to which the oscillations on the other wing are in an opposite direction to that on the first arm.

To further expound on the tunneling dynamics, we present the Husimi distribution of the evolving wave packet at specific instances of time in Fig.~\ref{fig:10}. The distribution first shows the initial Gaussian wave packet superposing with the vibrational curves above the separatrix in the pendulum's phase space. The wave packet evolves along the vibrational curves, remaining localized initially, and then transitions into a mixed shape with significant k-space spread, indicating dephasing (see the movie in the Supplemental Material ~\cite{NewRefSM}). These features can be exemplified, here, with the wave packet distribution at \(t = 0.25T_\text{tun}\), where the wave packet is shown extended along the vibrational curves. Here, the tunneling to the opposite wing also becomes apparent. At \(t = 0.50T_\text{tun}\), corresponding to the half-time between a tunneling event, the distribution shows nearly equal density on both spatial sides of the phase space. The localized structure in $k$-space at this time reflects coherent dynamics. The Husimi distribution shown for \(t = 1.00T_\text{tun}\) exhibits a complete transfer of the wave packet to the other side. The extension along momentum-space highlights an imperfect revival of the coherent dynamics.

Seen from the pendulum perspective, one can realize that the Bloch-like oscillations around a certain position correspond to the oscillations of the pendulum momentum when accelerating and decelerating during full rotations, and that the Wannier-Stark-like localization in one of the arms of the parabola corresponds to the momentum preserving its sign during full rotations. Thus, dynamics in the presence of tunneling appear as quantum beating between a clockwise rotating and a counterclockwise rotating pendulum.

\section{conclusion}
In conclusion, our study examines the eigenstates in a parabolic lattice system with a focus on near-separatrix dynamics. The analogy of states to closed and open curves of classical pendulum phase space highlights the nature of these dynamics, drawing parallels between classical and quantum behavior. While Bloch oscillations and dipole oscillations have been understood in the context of pendulum dynamics, our results reveal that their superposition within parabolic lattices can give rise to highly non-classical dynamics, which could be harnessed for generating non-classical states. Furthermore, our investigation demonstrates that the momentum of the initial wave packet plays a crucial role in dictating the system's dynamics, which can be tuned to realize various dynamical regimes, thereby broadening the scope of controllable quantum phenomena in such systems. We also emphasize the potential of enhancing the tunneling splitting between almost degenerate states well above the sepa\-ratrix through the use of bichromatic lattices, which could induce long-range tunneling dynamics at an experimentally relevant time scale. This approach may offer a pathway to observe long-range dynamical tunneling in binary parabolic optical lattices. Additionally, time-periodic driving may be employed to tune the energy splitting, and with judiciously chosen parameters one could induce dynamical tunneling between distant regular islands in a partly chaotic phase space. Exploring this, along with the effects of atom-atom interactions on tunneling dynamics, presents an exciting direction for future work. Thus, our findings suggest that the parabolic lattice system is well-suited for studying quantum dynamics, separatrix-like conditions, and long-range dynamical tunneling of macroscopic wave packets, providing a solid foundation for future experimental investigations.
\section*{Acknowledgments}
\par{U.A.~thanks the Deutscher Akademischer Austauschdienst (DAAD) for support through a doctoral research grant and Paderborn University for support via the Doctoral Scholarship 2024. M.H.~has been supported by the Deutsche Forschungsgemeinschaft (DFG, German Research Foundation) through Project No.~397122187.
We acknowledge support for the publication costs by the Open Access Publication Fund of Paderborn University.}

\end{document}